\documentclass[prb,twocolumn,showpacs,preprintnumbers]{revtex4}

\usepackage{graphicx}
\usepackage{dcolumn}
\usepackage{bm}

\begin{document}

\preprint{to be published in Phys. Rev. B}

\title{Competing types of quantum oscillations \\ in the 2D organic conductor
(BEDT-TTF)$_8$Hg$_4$Cl$_{12}$(C$_6$H$_5$Cl)$_2$
}

\author{Cyril Proust$^{1,2}$, Alain Audouard$^{1,2\dagger}$, Luc Brossard$^{1,2}$, Sergei~Pesotskii$^3$, Rustem Lyubovskii$^3$ and Rimma Lyubovskaya$^3$}

\affiliation {$^1$ Laboratoire de Physique de la Mati\`{e}re
Condens\'{e}e (UMR CNRS-UPS-INSA 5830), 135 avenue de Rangueil,
31077 Toulouse, France}

\affiliation {$^2$ Laboratoire National des Champs Magn\'{e}tiques
Puls\'{e}s (UMS CNRS-UPS-INSA 5642), 143 avenue de Rangueil, 31432
Toulouse, France}

\affiliation {$^3$ Institute of Problems of Chemical Physics,
Russian Academy of Sciences, Chernogolovka 142432, Russia}

\date{\today}

\begin{abstract}
Interlayer magnetoconductance of the quasi-two dimensional organic
metal (BEDT-TTF)$_8$Hg$_4$Cl$_{12}$(C$_6$H$_5$Cl)$_2$ has been
investigated in pulsed magnetic fields extending up to 36 T and in
the temperature range from 1.6 to 15 K. A complex oscillatory
spectrum, built on linear combinations of three basic frequencies
only is observed. These basic frequencies arise from the
compensated closed hole and electron orbits and from the two
orbits located in between. The field and temperature dependencies
of the amplitude of the various oscillation series are studied
within the framework of the coupled orbits model of Falicov and
Stachowiak. This analysis reveals that these series result from
the contribution of either conventional Shubnikov-de Haas effect
(SdH) or quantum interference (QI), both of them being induced by
magnetic breakthrough. Nevertheless, discrepancies between
experimental and calculated parameters indicate that these
phenomena alone cannot account for all of the data. Due to its low
effective mass, one of the QI oscillation series - which
corresponds to the whole first Brillouin zone area - is clearly
observed up to 13 K.
\end{abstract}


\maketitle

\section{\label{sec:level1}Introduction}

In some organic conductors, the Fermi surface (FS) presents quasi
2D pieces connected with small enough gaps, which offers the very
attractive opportunity to investigate still interesting questions
of fermiology such that the "competing coexistence" between
different types of quantum oscillations. This is the case of e. g.
the quasi-2D charge transfer salt $\kappa$-(ET)$_2$Cu(NCS)$_2$,
where ET stands for the donor molecule BEDT-TTF
(bisethylenedithia-tetrathiofulvalene). Indeed, the FS of this
compound appears to be adequately described by the textbook model
of a chain of coupled orbits introduced by Pippard \cite{1}. In
this compound, conventional Shubnikov de Haas (SdH) effect
resulting from magnetic flux quantization inside semiclassical
closed orbits and quantum interference (QI) \cite{2,3} between
open electron paths, which can be both induced by magnetic
breakthrough (MB) \cite{1,4,5} can occur. These phenomena yield
frequencies resulting from linear combinations of the two basic
frequencies a and b as observed in the oscillatory spectrum of the
magnetization \cite{6} and of the longitudinal magnetoresistance
\cite{7,8} although some of these combinations are forbidden in
the semiclassical picture. Similar frequency mixing induced by
oscillation of the chemical potential of a 2D-electron gas may
also contribute to \cite{7} or account for \cite{9} oscillatory
data. More recently, numerical computation of the de Haas-van
Alphen (dHvA) oscillation spectrum has been achieved. Based on a
realistic tight-binding model of quasi-2D organic conductors,
these computations also evidenced, although at quite high B/T
ratios, significant frequency mixing including the forbidden
frequencies. They occur as well for a fixed number of particles as
for a fixed chemical potential and are due to the field-dependent
interplay of electronic states from the different bands crossing
the Fermi level \cite{10}. The main still open question lies in
the relative weight of these different contributions to the data.

\begin{figure}
\centering
\resizebox{\columnwidth}{!}{\includegraphics*{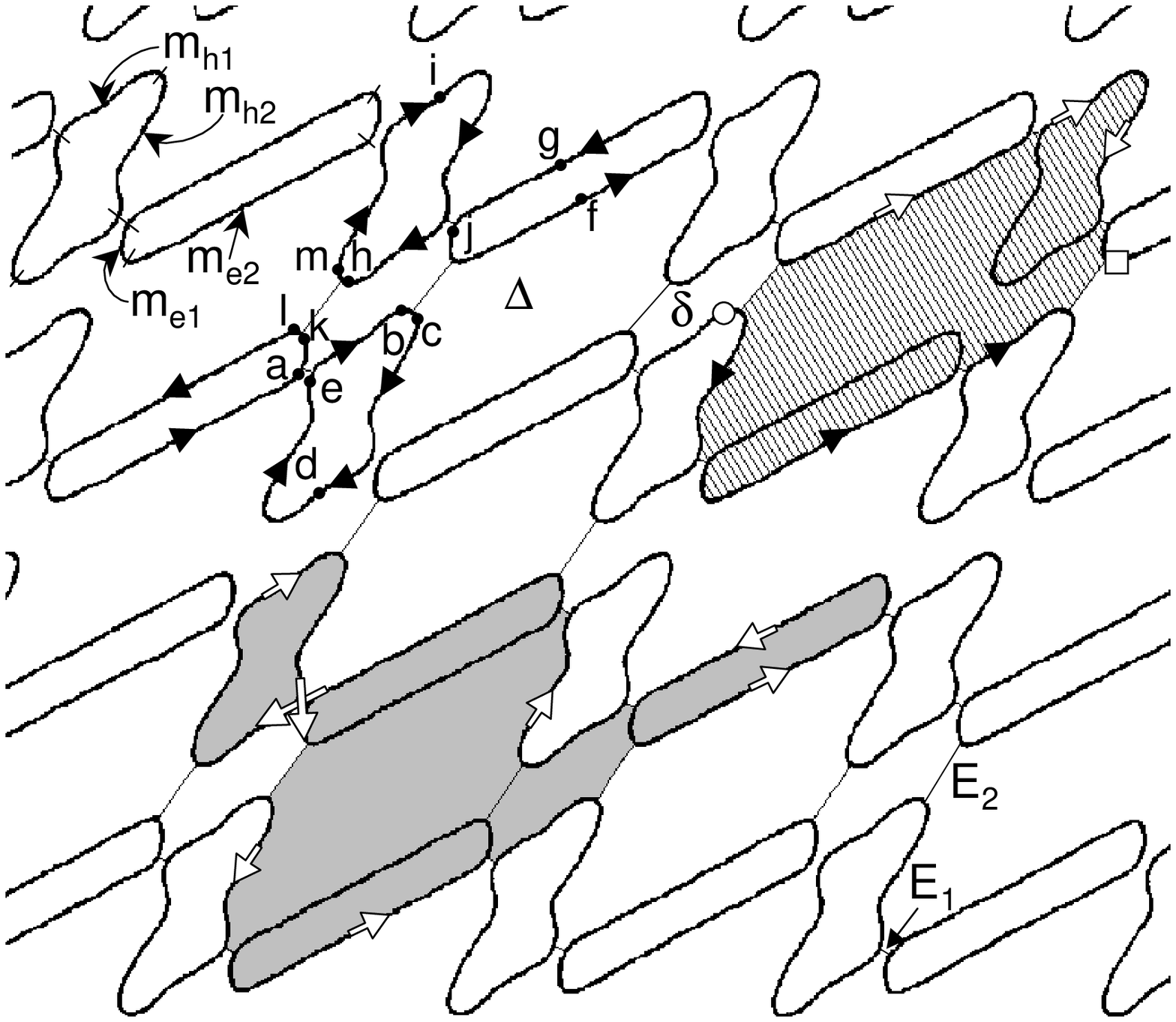}}
\caption{\label{Fig1} Fermi surface of (ET)$_8$Hg$_4$Cl$_{12}$(C$_6$H$_5$Cl)$_2$ according to band structure calculations of Vieros et al.
\cite{13}. Latin types are labels discussed in the text. Besides the $\delta$ and $\Delta$ orbits, the elongated electron and the hole closed
orbits can be observed. m$_{e1}$ (m$_{h1}$) and m$_{e2}$ (m$_{h2}$) are the weight factors, i. e. the partial effective mass linked to the short
and long part of the electron (hole) closed orbit, respectively. Shaded and hatched areas depict one of the SdH orbit and one of the two-arm
interferometers, respectively that can account for the b oscillation series. Open circles and open squares mark the extremities of the two-arm
interferometers.}
\end{figure}

The room temperature FS of (ET)$_8$Hg$_4$Cl$_{12}$(C$_6$H$_5$Cl)$_2$ results from the hybridization of two pairs of hidden quasi-1D sheets
\cite{11}, parallel to the (a*, b* + c*) and (a*, c*) planes \cite{12,13}. The resultant FS, obtained after raising of degeneracy is built up
with one hole and one elongated electron tube (see Figure 1). Although the cross section area of both electron and hole tubes amounts to 13
percent  of the FBZ area \cite{13}, the resulting orbits do not share the same topology and are separated from each other by two unequal gaps
labeled E$_1$ and E$_2$ in Figure 1. Provided these gaps are not too large, MB between electron and hole orbits can occur in magnetic field,
leading to a two-dimensional network of coupled orbits. This may give rise, besides quantum oscillations linked to the electron and hole closed
orbits, to additional oscillation frequencies that can be accounted for either by the semiclassical model of Falicov and Stachowiak \cite{4} or
by QI. Regarding the FS topology at low temperature, it is worth to notice that a metallic groundstate is stabilized in
(ET)$_8$Hg$_4$Cl$_{12}$(C$_6$H$_5$Cl)$_2$. Indeed, the conductivity exhibits a metallic behavior down to the lowest temperatures with a residual
resistivity ratio equal to \~ 100 and without any sign of (even imperfect) nesting of neither electron nor hole tubes \cite{12}.

Previous magnetoresistance experiments performed up to 15 teslas on (ET)$_8$Hg$_4$Cl$_{12}$(C$_6$H$_5$Cl)$_2$ crystals with the current injected
within the conducting bc-plane (in-plane configuration) \cite{14} exhibit one Shubnikov-de Haas (SdH) oscillation series, referred to as the a
series hereafter, with a frequency F$_a$ = 250 T corresponding to a cross section of $\approx$ 11 \% of the FBZ area \cite{14}. Nevertheless,
when the current is injected in the direction a*, normal to the conducting plane (interlayer configuration), a complex oscillatory behavior is
observed, in particular at high magnetic field \cite{15}. Namely, in addition to a frequency F$_b$ = 2200 T, corresponding to $\approx$ 100 \%
of the FBZ area, other frequencies which are linear combinations of the frequencies F$_a$ = 240 T and F$_\delta$ = 150 T (7 \% of the FBZ area),
respectively have been observed.

The aim of this paper is to show that the oscillatory behavior of the interlayer magnetoresistance of the
(ET)$_8$Hg$_4$Cl$_{12}$(C$_6$H$_5$Cl)$_2$ organic conductor results from several contributions, including MB-induced QI effects. This will be
achieved through the analysis of the temperature and field magnitude and orientation dependencies of the oscillation spectrum.

\section{\label{Exp}Experimental}

The studied crystal was a platelet with approximate dimensions (1$\times$1$\times$0.1)~mm$^3$, the largest faces being parallel to the
conducting bc-plane. Electrical contacts were made to the crystal using annealed gold wires of $20$~mm in diameter glued with graphite paste.
Alternating current ($400$~mA, $50$~kHz) was injected parallel to the a* direction (interlayer configuration). A lock-in amplifier with a time
constant of 100 ms was used to detect the signal across the potential leads. If the measurements, performed during the decay of the pulsed
fields of the LNCMP ($36$~T, $1.2$~sec) were noiseless, this should allow to derive reliable oscillatory data down to e. g. $11$~T and $1$~T for
a frequency of $2200$~T and $150$~T, respectively (see Ref. \cite{16}).

Data analysis is based on Fourier transforms (FT) calculated with an elevated cosine window in a given field range from B$_{min}$ to B$_{max}$.
In the following, the amplitude of a given oscillation series at the mean field value B = 2 / (1/B$_{min}$ + 1/B$_{max}$) is determined by the
ratio of the amplitude of the FT to (1/B$_{min}$ - 1/B$_{max}$). The orientation of the magnetic field is defined by the angle $\theta$ between
the field direction and the normal to the conducting bc-plane. The sign of $\theta$ is arbitrary.

\section{\label{Res}Results and discussion}

\subsection{\label{Osc}Oscillatory spectrum}

\begin{figure}
\centering \resizebox{\columnwidth}{!}{\includegraphics*{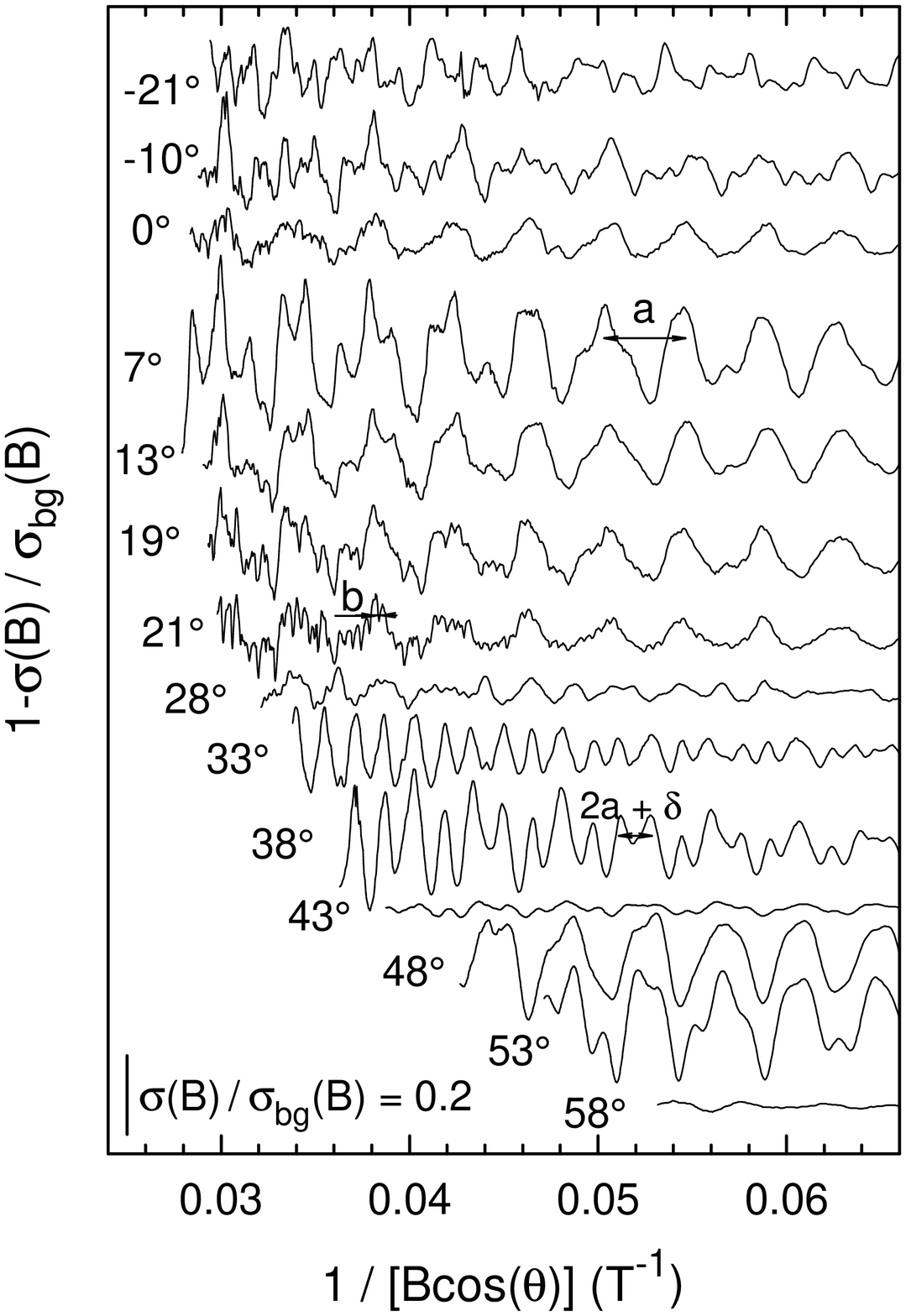}}
\caption{\label{Fig2} Oscillatory part of the magnetoconductance at $\sim$ $1.7$~K for different orientations of the magnetic field.
$\sigma_{bg}$ is the field-dependent background part of the conductivity. $\theta$ is the angle between the magnetic field direction and the
normal to the conducting plane.}
\end{figure}

\begin{figure}
\centering \resizebox{\columnwidth}{!}{\includegraphics*{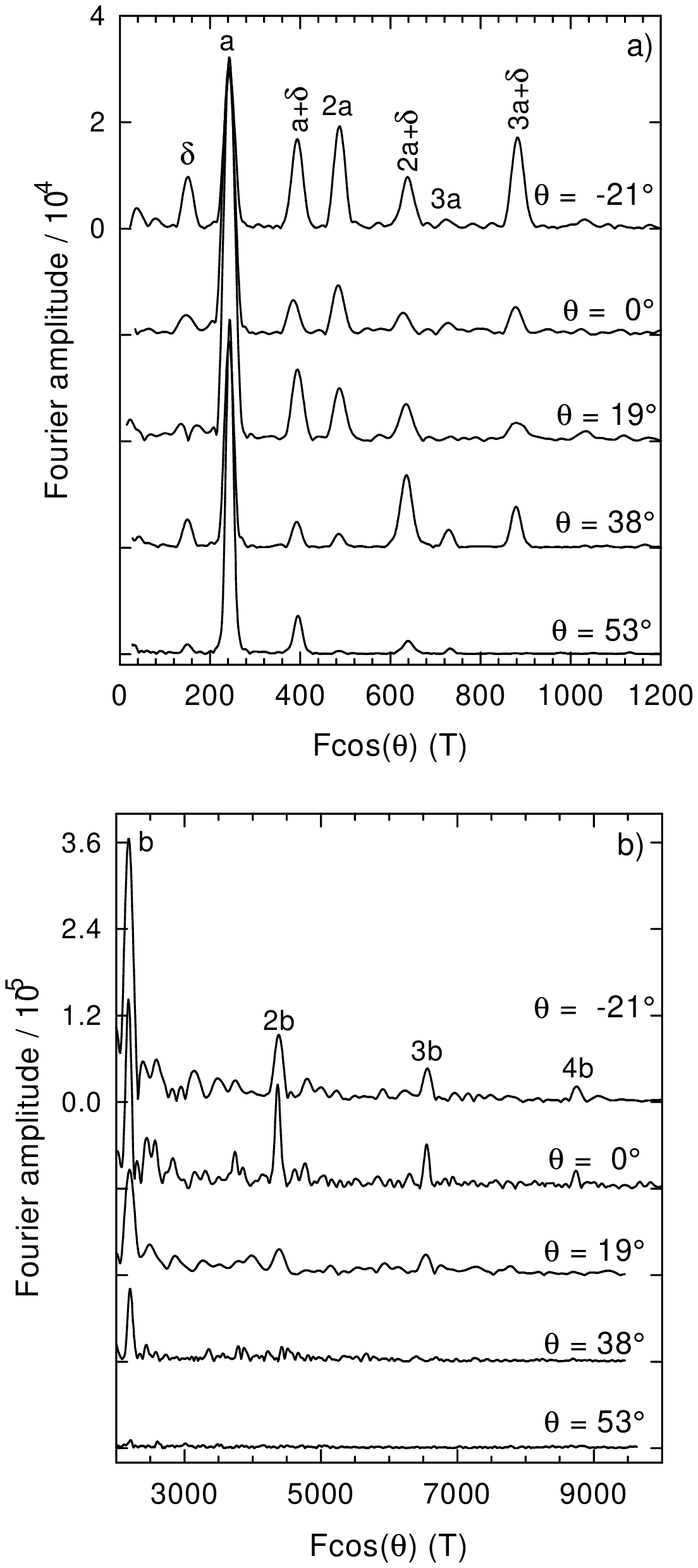}}
\caption{\label{Fig3} Fourier transforms deduced from data in Figure 2. The magnetic field window is 10 - 30 T and 20 - 35.2 T for Fig. 3a and
Fig. 3b, respectively.}
\end{figure}

\begin{figure}
\centering \resizebox{\columnwidth}{!}{\includegraphics*{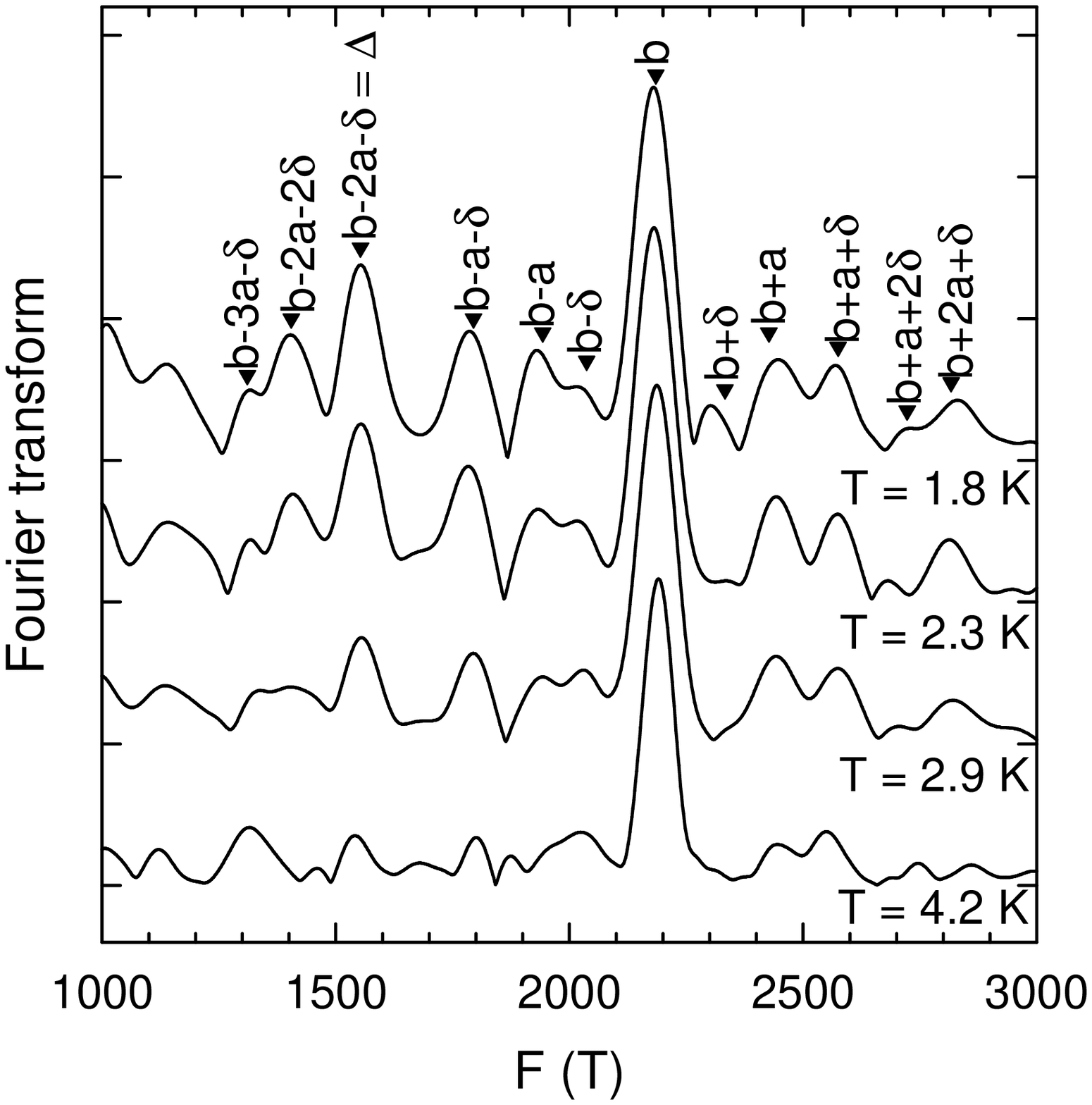}}
\caption{\label{Fig4} Fourier transforms deduced from data at
different temperatures for $\theta$ = 0$^{\circ}$ in the magnetic
field range 18 - 35.7 T. Frequency combinations involving a, b and
$\delta$ orbits are indicated in the figure. Full triangles are
marks calculated with F$_a$ = 241.5~T, F$_{\delta}$ = 149~T and
F$_b$ = 2185~T. Note that the oscillation series b-2a-$\delta$
corresponds to the $\Delta$ orbit.}
\end{figure}

Fig.~\ref{Fig2} displays the oscillatory magnetoconductance for
different orientations of the magnetic field at a temperature of
$\sim$ $1.7$~K. FT deduced from data in Fig.~\ref{Fig2} are
displayed in Fig.~\ref{Fig3}. A complex oscillatory behavior is
observed since up to 6 fundamental frequencies - without counting
some harmonics - are displayed in Fig.~\ref{Fig3}. Moreover, these
6 frequencies follow the orbital behavior expected for a
two-dimensional FS in the explored angle range from -21$^{\circ}$
to +72$^{\circ}$, as demonstrated in Fig.~\ref{Fig3} for some
angles. In the field range between 10 and 30 teslas (see
Figure~3a), the observed frequencies can be regarded as linear
combinations of F$_a$ and F$_{\delta}$ with F$_a$ = (241.5 $\pm$
2.0)~T and F$_{\delta}$ = (149 $\pm$ 2)~T. In addition, an
oscillation series with a frequency F$_b$ = (2185 $\pm$ 15)~T and
up to 3 harmonics are observed in the higher field range between
20 and 35.2 T (see Figure 3b). These three frequencies correspond
to cross section area of 11.0 $\pm$ 0.1, 6.8 $\pm$ 0.1 and 100
$\pm$ 1 percent of the FBZ area, respectively. According to band
structure calculations, the cross section area of the electron and
the hole orbits corresponds to approximately 13 \% of the FBZ area
\cite{13}. Keeping in mind that a discrepancy of few percents
between experimental data and band structure calculations is an
usual feature, it can be assumed that the calculated FS is in
qualitative agreement with the experimental data. Hence F$_{a}$ is
associated to the closed electron and hole orbits (referred to as
the a orbits in the following) and F$_{\delta}$ to the $\delta$
orbit. F$_b$, which corresponds to the whole FBZ area, accounts
for an orbit that involves 2a + $\Delta$ + $\delta$. Owing to the
experimental values of the frequencies F$_a$ and F$_{\delta}$, the
cross section of the $\Delta$ orbit amounts to 71 \% of the FBZ
area. Finally, it is important to notice that more than ten
frequency combinations involving F$_b$ are also observed in FT
performed in the high field range (see Fig.~\ref{Fig4}). In
particular, the frequency linked to b-2a-$\delta$, which
corresponds to the $\Delta$ orbit is clearly evidenced in the
figure. In order to assign the above reported frequencies to
k-space SdH orbits or QI paths, it is important to keep in mind
that, following Falicov and Stachowiak \cite{4} and Shoenberg
\cite{5}, areas enclosed by electron and hole parts of a MB orbit
bears opposite sign. One of the possible consequence is that very
large SdH orbits and QI paths including both electron and hole
parts may account for a given (even rather low) frequency,
although with a reduced damping factor and a large effective mass,
as discussed in the next section. For example, F$_{\delta}$ can be
attributed, among others, to both the semiclassical MB closed
orbit (a b c d e b f g h m i h l a) and the QI path (a k l)-(a b c
d e b f g h l) (see Fig.~\ref{Fig1}). It is worth to note that
magnetic interaction MI can also induce frequency combinations in
dHvA oscillations spectrum when the FS is composed of several
orbits. However, recent measurements \cite{17} have revealed that,
as it is usually the case for most organic conductors, the value
of the magnetization of the isostructural compound
(ET)$_8$Hg$_4$Cl$_{12}$(C$_6$H$_5$Cl)$_2$ remains rather weak even
at high magnetic field. This latter result makes unlikely a
significant contribution of MI to the oscillation spectrum. In the
following, we will examine the possible contribution of QI and
conventional SdH effect to the observed oscillatory behavior
through the temperature and magnetic field dependencies of the
oscillation amplitude. Unfortunately, the oscillation series
resulting from frequency combinations involving the F$_b$
frequency (see Fig.~\ref{Fig4}) will not be considered due to too
small amplitude and (or) too steep field and temperature
dependencies.

\subsection{\label{Cal}Calculation of effective masses and damping factors}

According to the conventional Lifshitz-Kosevich (LK) model, the
field and temperature dependence of the oscillatory part of the
conductivity can be accounted for by:

\begin{equation}
1-\frac{\sigma(B)}{\sigma_{{bg}}(B)}=\sum_{{i}}A_i\cos \left[ 2\pi\left(\frac{F_i}{B}-\gamma_i\right) \right]
\end{equation}

where $\sigma_{bg}$(B) is the field-dependent monotonous part of
the conductivity. i stands for the indices of the oscillation
series and $\gamma_i$ is the Onsager's phase factor. Harmonics
contribution can be included in the equation. Neglecting the spin
splitting damping term, the oscillation amplitude A$_i$ is given
by:

\begin{equation}
A_{{i}} \propto R_{{T}}(i)R_{{D}}(i)R_{{MB}}(i)
\end{equation}

where R$_T$ and R$_D$ are the temperature and Dingle damping terms, respectively. R$_{MB}$, which will be considered latter on, is the damping
term which accounts for the contribution of MB. As usual, the Dingle damping term is expressed as:

\begin{equation}
R_{{D}}(i) =
exp\left[-\frac{u_{{0}}T_{{D}}(i)m_{{c}}(i)}{B}\right]
\end{equation}

where (u$_0$ = 14.694~T/K); T$_D$(i) and m$_c$(i) are the Dingle temperature and the effective cyclotron mass, respectively. R$_T$(i) is given
by:

\begin{equation}
R_{{T}}(i) \propto \frac{Tm_{{c}}(i)/B^n}{\sinh \left[-u_{{0}}Tm_{{c}}(i)/B \right]}
\end{equation}

In the two- and three-dimensional case, n is equal to 1 and 1/2, respectively \cite{5}.

Following Falicov and Stachowiak \cite{4}, the effective mass linked to electron and hole orbits can be expressed as m*$_e$ = 2(m$_{e1}$ +
m$_{e2}$) and m*$_h$ = 2(m$_{h1}$ + m$_{h2}$), respectively. The weight factors m$_{e1}$, m$_{e2}$, m$_{h1}$ and m$_{h2}$, which can be regarded
as absolute values of partial cyclotron mass parameters, are defined in Fig.~\ref{Fig1}. Since the oscillation series with frequency F$_a$
results from the contribution of the electron and hole orbits, the resultant effective cyclotron mass has been assumed equal to m*$_a$ = (m*$_e$
+ m*$_h$) / 2, i.e. m*$_a$ = m$_{e1}$ + m$_{e2}$ + m$_{h1}$ + m$_{h2}$. Same type of calculation has been performed for the other SdH orbits as
reported in Table 1.

\begin{table*}
\caption{\label{tab:table1}Experimental data and calculated
parameters relevant to the observed oscillation series. F($\theta$
= 0$^{\circ}$) is the oscillation frequency deduced from
experimental data for the magnetic field applied perpendicular to
the conducting plane. m* and m$_c$ are the calculated effective
mass and the experimental effective cyclotron mass deduced from
the conventional LK model (in the temperature range below 8~K for
b and $\delta$ oscillations), respectively relevant to the
considered oscillation series. The field-dependent part of the
damping factors K$_{SdH}$ and K$_{QI}$ are defined in Eq. (5) and
(6), respectively. Only the SdH orbits and QI paths yielding the
highest damping factors are considered in the table.}
\begin{tabular}{|c|c|c|c|c|c|c|c|}

\hline
 &\multicolumn{3}{c|}{Experimental data}&\multicolumn{4}{c|}{Calculations}\\
\cline{2-8}
 & & & & \multicolumn{2}{c|}{SdH oscillations}&\multicolumn{2}{c|}{Quantum interference oscillations}\\
\cline{5-8}
orbit&F($\theta=0^\circ$)&m$_c$&m$_c$/m$_c$(a)&m*/m*(a)&K$_{SdH}$&m*/m*(a)&K$_{QI}$\\
 \hline
$\delta$&149 $\pm$ 2&0.50 $\pm$ 0.15&0.43 $\pm$ 0.18&4&q$_1^6$q$_2^4$p$_1^2$p$_2^2$&2&$q_1^4q_2^4p_1^2p_2^2$\\
\hline
a&241.5 $\pm$ 2&1.17 $\pm$ 0.13&1&1&$q_1^2q_2^2$&not relevant&not relevant\\
\hline
a+$\delta$&391 $\pm$ 4&1.02 $\pm$ 0.08&0.87 $\pm$ 0.17&3&$q_1^4q_2^4p_1^2p_2^2$&1&$q_1^2q_2^2p_1^2p_2^2$\\
\hline
2a+$\delta$&633 $\pm$ 4&1.95 $\pm$ 0.10&1.67 $\pm$ 0.27&2&$q_1^2q_2^2p_1^2p_2^2$&2$\left |m_{e2}-m_{h2}\right |/m*(a)$&$q_1^2q_2^2p_1^2p_2^2$ or $q_1^4q_2^2p_1^2p_2^2$\\
\hline
3a+$\delta$&875 $\pm$ 15&0.73 $\pm$ 0.15&0.62 $\pm$ 0.20&3&$q_1^4q_2^4p_1^2p_2^2$&1&$q_1^4q_2^4p_1^2p_2^2$\\
\hline
b&2185 $\pm$ 15&0.5 $\pm$ 0.1&0.43 $\pm$ 0.13&4&$q_1^6q_2^2p_1^2p_2^6$&0&$q_2^4p_1^4p_2^2$ or $q_1^4p_1^2p_2^4$\\
\hline

\end{tabular}
\end{table*}

In the framework of the QI model \cite{2,3}, the effective mass is given by the energy derivative of the phase difference ($\varphi_i$ -
$\varphi_j$) between the two different routes i and j of a two-arm interferometer. Within this model, $\partial(\varphi_i - \varphi_j) /
\partial \epsilon = heB~
\partial S_k / \partial\epsilon$, where S$_k$ is the reciprocal space area bounded between the two arms. Since $\partial(\varphi_i - \varphi_j) /
\partial \epsilon$ is proportional to the difference between the effective mass of the two arms of the interferometer, the associated effective
mass is given by m* = $\mid$ m*$_i$ - m*$_j$ $\mid$ where m*$_i$ and m*$_j$ are the partial effective masses of the routes i and j. The
calculated values for the QI orbits are given in Table 1. It should be kept in mind that a given oscillation series can be accounted for by
several types of QI paths or SdH orbits with different damping factors. Data in Table 1 is restricted to orbits yielding the highest damping
factors i. e. with the lowest number of MB junctions and the lowest effective mass.

For a given oscillation series, noticeable differences between effective masses linked to either SdH or QI can be observed. E. g. m*$_{2a +
\delta}$ is equal to 2 m*$_a$ in the case of SdH while m*$_{2a + \delta}$ is equal to 2$\mid m_{e2}-m_{h2}\mid$ in the case of QI which is
certainly much lower than m*$_a$.

We discuss now the damping factor R$_{MB}$ entering Eq. (2). According to Falicov and Stachowiak \cite{4}, the damping factor for a SdH orbit
can be written as:

\begin{equation}
R_{MB}^{SdH}(i)=\prod_{g=1,2} p_g^{n_{pg}}q_g^{n_{qg}} \exp \left[i\left(n_{pg}\varphi_p+n_{qg}\varphi_q\right)\right]
\end{equation}

The indices g stand for the two different gaps between electron
and hole orbits (see Fig.~\ref{Fig1}). $\varphi_p$ and $\varphi_q$
are phase factors $(\varphi_p + \varphi_q = \pm \pi/2)$. The
integers n$_{pg}$ and n$_{qg}$ are respectively equal to the
number of MB and Bragg reflections encountered along the path of
the quasiparticle. The MB and Bragg reflection probabilities are
given by $p_g^2=exp(-B_g/B)$ and $\mid q_g \mid^2=1-p_g^2$ ,
respectively where B$_g$ is the gap-dependent MB field. In the
following, the field-dependent part of $R_{MB}^{SdH}(i)$  is
expressed as K$_{SdH}(i)=\prod_{g=1,2} p_g^{n_{pg}}q_g^{n_{qg}}$.


In the case of a two-arm interferometer, the damping term for QI is given, according to Harrison et al. \cite{7,18} by:

\begin{equation}{\label{Eq7}}
  R_{MB}^{QI}(i) \propto \prod_{k,g=1,2}
p_{kg}^{n_{pg}}q_{kg}^{n_{qg}}
\exp\left(-\frac{\pi}{\omega_i\tau_t}\right)
\end{equation}{\label{Eq7}

In this expression, i and g indices have the same meaning as in
Eq. (5) while k indices stands for each of the two arms of the
interferometer. n$_{pg}$ and n$_{qg}$ are the number of MB and
Bragg reflections, respectively encountered by each of the arms of
the interferometer. As stated in \cite{18}, the relevant effective
mass m$_i$ entering $\omega_i=\hbar eB/m_i$  is the sum of the
partial effective masses of the two branches of the
interferometer. In addition, as pointed out by Stark and Friedberg
\cite{3} and contrary to the scattering time involved in the
Dingle damping factor, which is usually assumed
temperature-independent, the quantum state lifetime $\tau_t$
includes the temperature-dependent electron-phonon interaction.
Assuming, that 1/$\tau_t$ is proportional to the zero field
resistivity $\rho_0$(T) yields:

\begin{equation}
R_{MB}^{QI}(i) \propto K_{QI}(i)\exp \left(-\alpha(i)\rho_0\right)
\end{equation}

where $\alpha$(i) and $\rho_0$ are field- and temperature-dependent, respectively.

As pointed out above and contrary to the case of e. g.
$\kappa$-(ET)$_2$Cu(NCS)$_2$, several QI paths or SdH orbits with
different topologies may contribute to a given oscillation series,
even restricting ourselves to the orbits or paths with the highest
damping factor. As an example, the a+$\delta$ series can be
accounted for by the QI paths with the arms (a k l) - (a b f g h
l) and (e b c) - (e k m i j c) (see Fig.~\ref{Fig1}).
Nevertheless, both of them include the same four MB and four Bragg
reflections involving the small and the large gap between electron
and hole orbit. This leads to the field-dependent part of the
damping factor K$_{QI}(a+\delta) = q_1^2q_2^2p_1^2p_2^2$ (see Eq.
(6) and (7)). In addition to SdH orbits which may present either
hole (c d e k m i j c orbit) or electron (a b f g h l a orbit)
character, the 2a+$\delta$ series can be accounted for by the two
interferometers with the arms (a k m i j) - (a b f g j) and (b f g
h m) - (b c d e k m). However, contrary to the case of the
a+$\delta$ series, these two QI orbits yield different damping
factors, namely $q_1^4q_2^2p_1^2p_2^2$ and $q_1^2q_2^4p_1^2p_2^2$,
respectively. Regarding the b oscillation series, it can be
accounted for by semiclassical MB orbits, such as the one marked
with shaded area in Fig.~\ref{Fig1}, with large effective mass and
reduced MB damping factors (see Table 1). A lot of QI paths can
also account for F$_b$, even without taking into account the
interferometers which involve QI paths with strongly different
arms length and bear reduced MB damping factors (12 MB junctions)
and large effective mass (m* = 2m*(a)). Among those with a zero
effective mass, some interferometers also involve a large number
of MB junctions. Since they are the most probable, only those
which involve 10 MB junctions (see the hatched area in
Fig.~\ref{Fig1}) are considered in Table 1.

\subsection{\label{Dat}Data analysis}

\begin{figure}
\centering \resizebox{\columnwidth}{!}{\includegraphics*{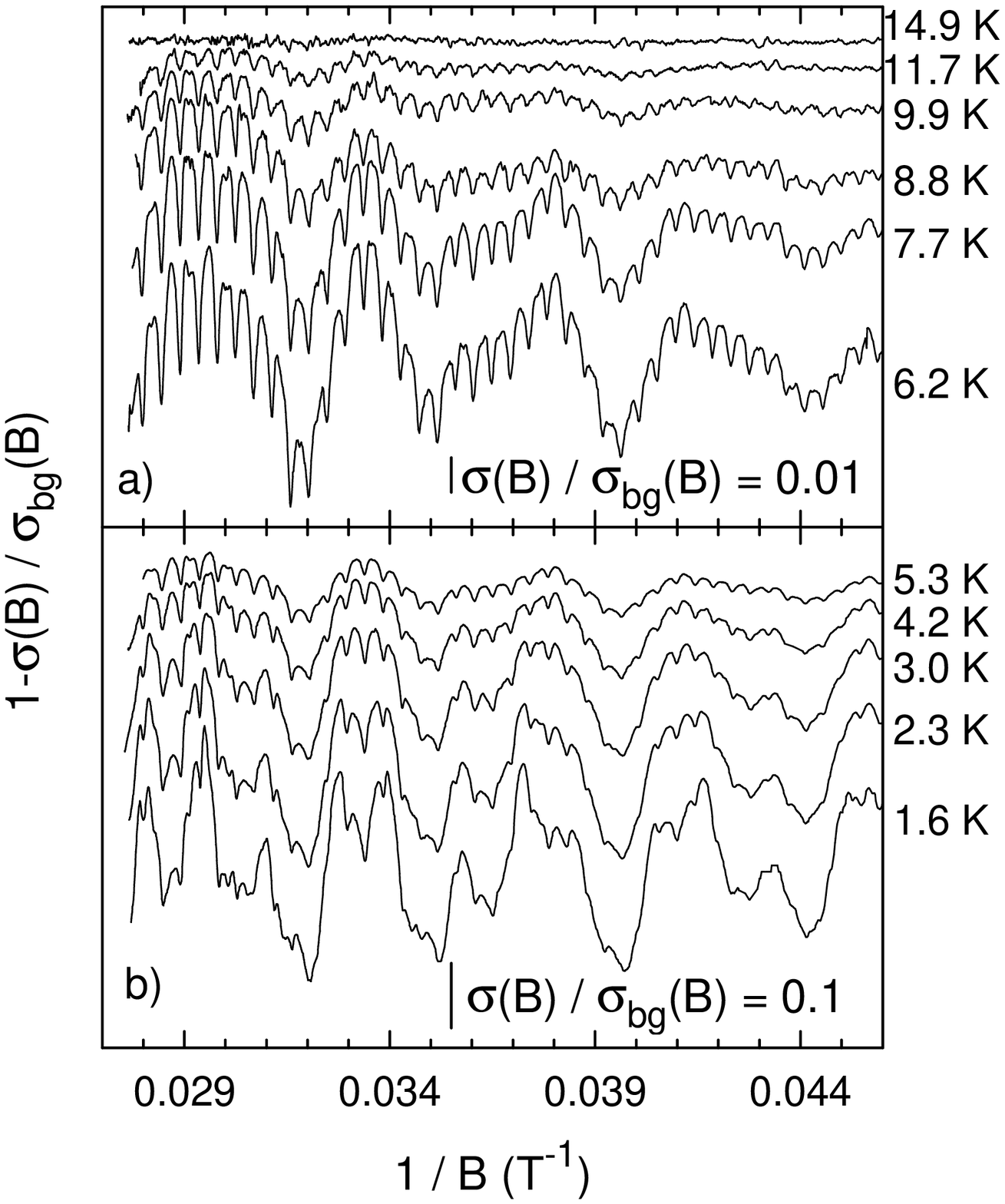}}
\caption{\label{Fig5} Oscillatory part of the magnetoconductance at different temperatures in the field range 21.7 to 36 T. The angle between
the magnetic field direction and the normal to the conducting plane is $\theta$ = -13$^{\circ}$.}
\end{figure}

\begin{figure}
\centering \resizebox{\columnwidth}{!}{\includegraphics*{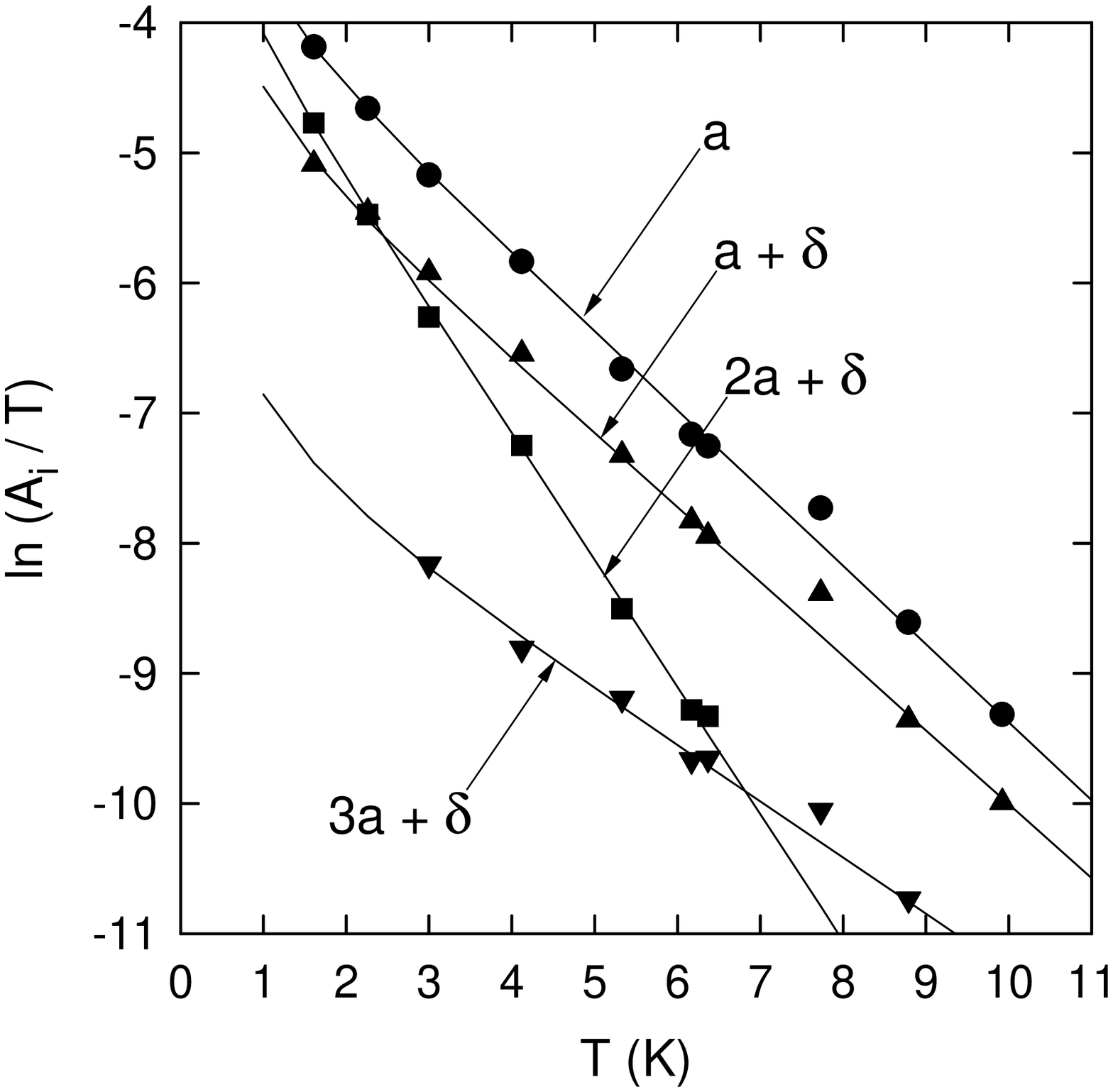}}
\caption{\label{Fig6} Temperature dependence of the oscillation amplitude of some of the series observed in Fig.~\ref{Fig5}. The angle between
the magnetic field direction and the normal to the conducting plane is $\theta$ = -13$^{\circ}$. The mean field value is 29.57 T, 28.29 T, 29.57
T and 27.66 T, respectively for the a, a+$\delta$, 2a+$\delta$ and 3a+$\delta$ series. Solid lines are best fits to Eq. (4).}
\end{figure}

The oscillatory part of the magnetoconductance is displayed in the field range 22 to 36 T for $\theta$ = -13$^{\circ}$ in Fig.~\ref{Fig5}. The
temperature dependence of the amplitude of the oscillations for the a, a+$\delta$, 2a+$\delta$ and 3a+$\delta$ series observed in the
oscillatory spectrum is displayed in Fig.~\ref{Fig6} in the temperature range up to 10 K. A good agreement with the conventional LK model is
observed as it is the case for the $\delta$ and b oscillations below $\sim$ 8K (see Fig.~\ref{Fig7}). However, Fig.~\ref{Fig7} displays strong
downward deviations from above $\sim$ 8 K for these latter series. These deviations from LK model are discussed later on. The cyclotron
effective masses have been determined for different directions and mean values of the magnetic field. In the main, slight variations of the
measured effective cyclotron mass parameter can be observed at high magnetic field, likely due to the strongly two-dimensional character of the
FS \cite{19}. Stronger downward deviations are nevertheless observed in some cases e. g. for m$_c$(a) at $\theta$ = 33$^{\circ}$ and
m$_c$(a+$\delta$) at $\theta$ = 22$^{\circ}$. The values deduced from experimental data are given in Table 1, assuming that reliable values of
the effective cyclotron mass are obtained at low magnetic field.

In the framework of the Fermi liquid theory, effective cyclotron
masses are renormalised by electron-phonon interactions and
electron correlation, accounted for by multiplicative factors (1 +
$\lambda_{e-ph}$) and (1 + $\lambda_{e-e}$), respectively where
$\lambda_{e-ph}$ and $\lambda_{e-e}$ are the strength of the
interactions. Assuming this model holds in the present case,
allows us to compare experimental values of m$_c$(i) / m$_c$(a) to
calculated values of m*$_i$ / m*$_a$.

\begin{figure}
\centering \resizebox{\columnwidth}{!}{\includegraphics*{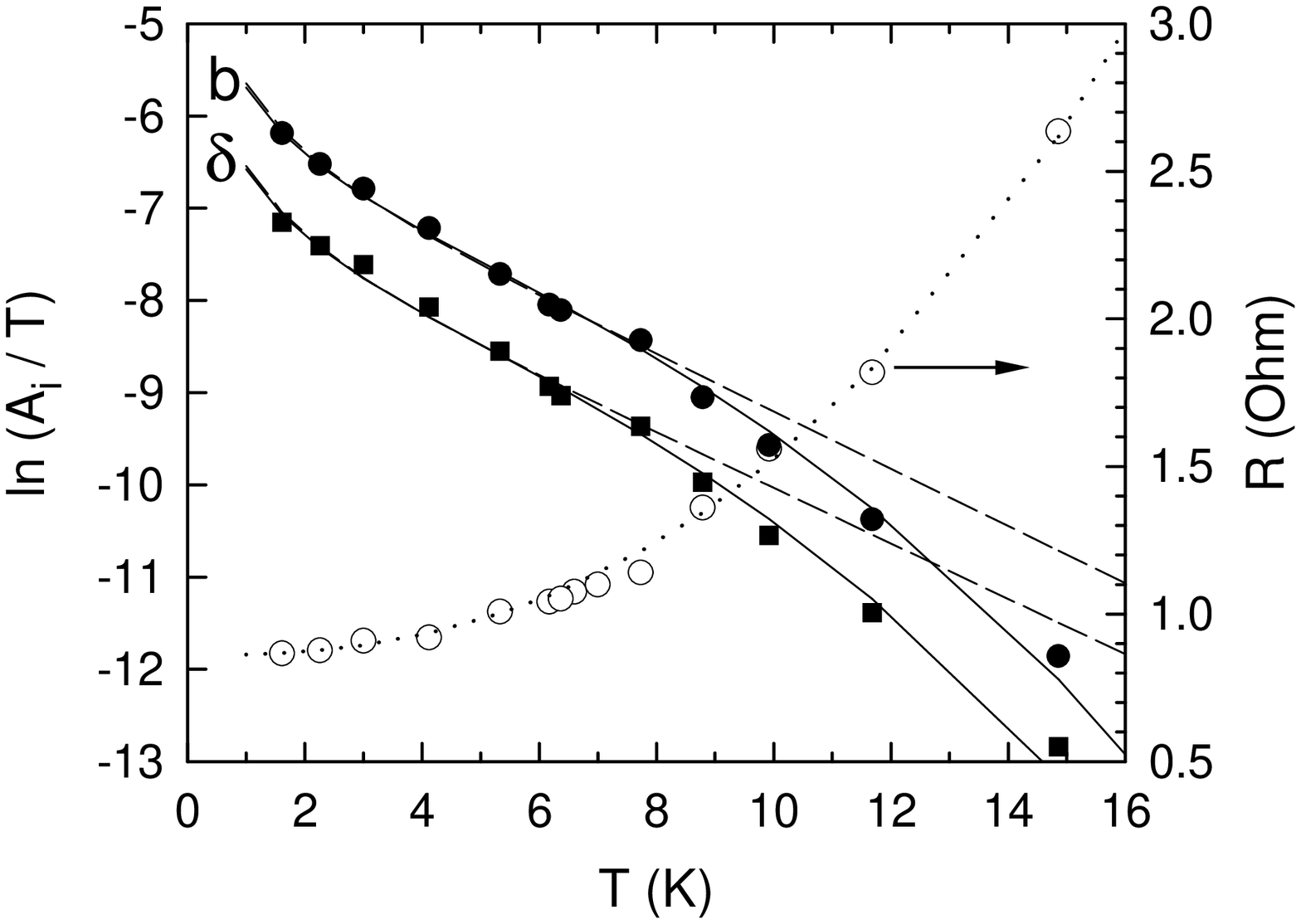}}
\caption{\label{Fig7} Temperature dependence of the amplitude of
the b (full circles) and $\delta$ (full squares) oscillations. The
mean field value is 22.66 T and 29.57 T, respectively for the
$\delta$ and b series. The angle between the magnetic field
direction and the normal to the conducting plane is $\theta$ =
-13$^{\circ}$. Dashed lines are best fits to the LK model (see Eq.
(4)) as done in Fig.~\ref{Fig6}. Full lines are best fits to the
LK model, assuming a zero effective cyclotron mass and taking into
account the temperature dependence of the quantum state lifetime
according to Eq. (7). The temperature dependence of the zero field
resistance (empty circles) is also displayed. Dotted line is a
guide for the eye and accounts for a T$^{2.46}$ dependence (see
text).}
\end{figure}

Effective cyclotron mass m$_c$(a+$\delta$) is close to m$_c$(a)
(see Table 1) which is in agreement with QI phenomenon. Similarly,
m$_c$(2a+$\delta$) value is in between m$_c$(a) and 2m$_c$(a)
which suggests a significant contribution of conventional SdH.
Oppositely, m$_c$($\delta$) and m$_c$(3a+$\delta$) are lower than
m$_c$(a) which invalidates both the conventional SdH and the QI
models. It must be pointed out that, the a+$\delta$ series is not
observed in dHvA experiments performed at low magnetic field
\cite{15}, although these dHvA data present a better
signal-to-noise ratio than the present conductivity data. This
result suggests that QI do contribute to the oscillatory behavior.
Otherwise, the 2a+$\delta$ frequency combination, which mainly
results from SdH, is still visible in the dHvA experiment.
Nevertheless, the frequencies linked to the $\delta$ and to the
3a+$\delta$ orbits which are not accounted for by neither
conventional SdH nor QI are also not detected in the dHvA data.
Owing to the large cross section of the b orbits, m$_c$(b) deduced
from the low temperature part of the data is low when compared to
m$_c$(a) since m$_c$(b) / m$_c$(a) $\sim$ 0.4 (see dashed lines in
Fig.~\ref{Fig7}). This rules out the conventional SdH model and
suggests that the b oscillation series may result from QI.
Nevertheless, only the interferometers with a zero effective mass
should significantly contribute to the oscillatory behavior. Such
a discrepancy has already been observed in the case of the 3D
LaB$_6$ compound which also exhibits a QI orbit with an extremal
cross section equal to the FBZ area and for which a zero effective
mass is predicted \cite{18}. This discrepancy can be accounted for
by Eq. (7) which assumes that the relevant lifetime arising in Eq.
(6) is proportional to the interlayer conductivity in zero-field.
This is demonstrated in Fig.~\ref{Fig7} where solid lines are best
fits to Eq. (2) assuming a zero effective cyclotron mass (i. e.
R$_T$(b) defined in Eq. (4) is temperature-independent) and a
temperature dependence of R$_{MB}^{QI}$ given by Eq. (7). Even
better agreement between data and Eq. (7) is obtained assuming
1/$\tau_t$ is proportional to T$^2$, which constitutes a signature
of the Fermi liquid behavior. It can be noticed in Fig.~\ref{Fig7}
that Eq. (7) also holds for the $\delta$ oscillation series
although there is no theoretical justification for this behavior
in the framework of the SdH and QI models since much higher values
of m$_c$($\delta$) are predicted in both cases (see Table 1).

\begin{figure}
\centering \resizebox{\columnwidth}{!}{\includegraphics*{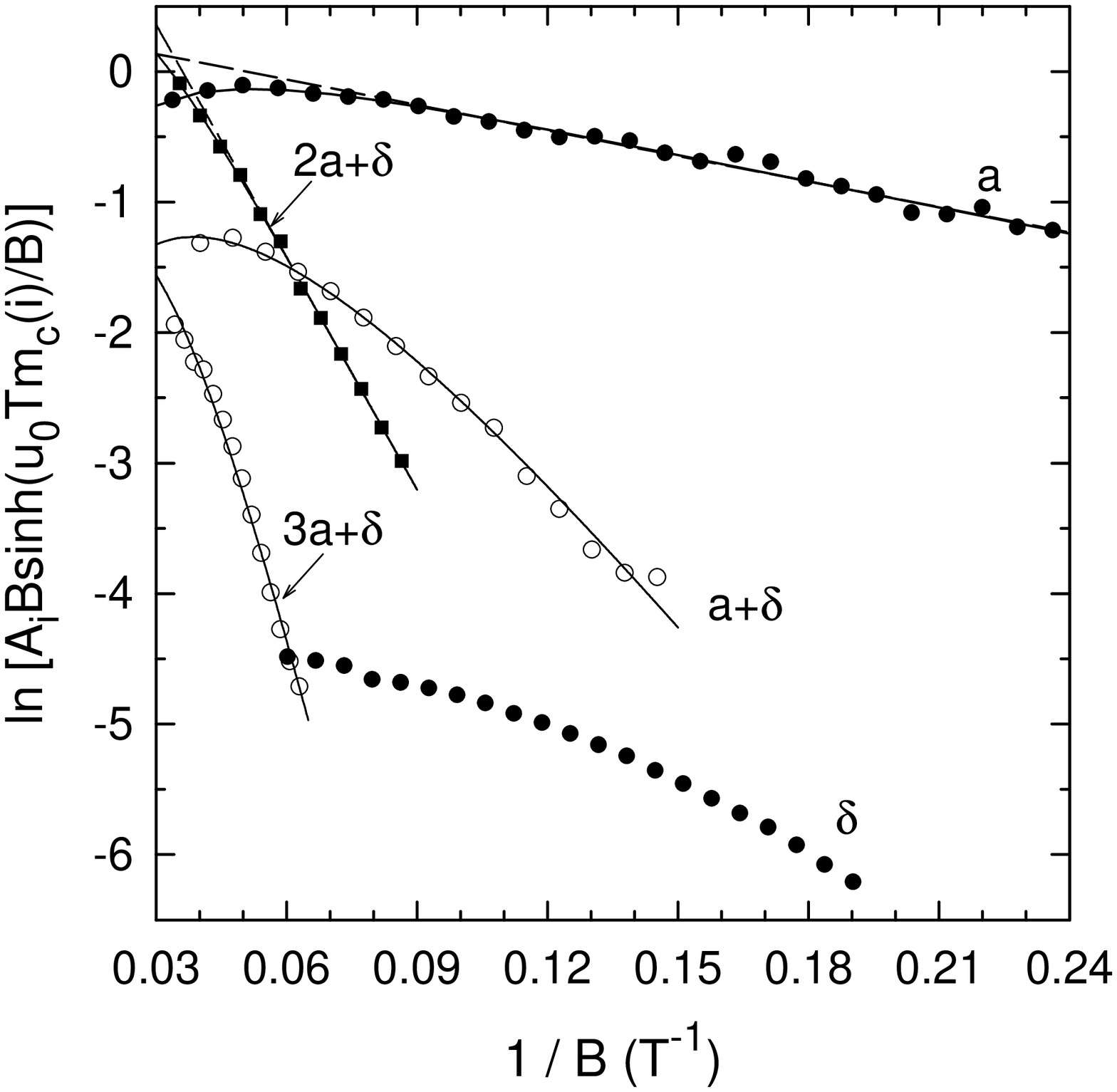}}
\caption{\label{Fig8} Dingle plots of the oscillation series observed in the data at 1.6 K and $\theta$ = -13$^{\circ}$ (see Fig.~\ref{Fig5}).
Solid and dashed lines are best fits to the data taking and without taking into account contribution of magnetic breakdown damping factor,
respectively (see text). The values of the effective cyclotron mass parameter m$_c$ are given in Table 1.}
\end{figure}

Additional information can be derived from the field dependence of the oscillation amplitude. Fig.~\ref{Fig8} displays conventional Dingle plots
of the various oscillation series at 1.6 K (see Fig.~\ref{Fig5}). Data at $\theta$ = -13$^{\circ}$ have been chosen since no significant field
dependence of the various effective cyclotron masses has been observed for this field direction. The b series has not been considered since
reliable data can be derived for the b oscillations in a very narrow field range only, likely due to their weak amplitude. The two-dimensional
case (n = 1 in Eq. (4)) is considered. Dashed lines in Fig.~\ref{Fig8} are best fits of Eq. (2) to the data without taking into account any
contribution of MB damping factor. Downward deviations from linearity are observed at high magnetic field \cite{20}. Solid lines in the figure
are best fits to the data including MB damping factor (see Table 1) relevant to either SdH (a and 2a+$\delta$ series) or QI (a+$\delta$ series).
It is worth to note that the same damping factor holds for SdH and QI in the case of the 3a+$\delta$ oscillation series. Generally speaking, a
very large uncertainty in the derived values of MB fields is obtained. Assuming same MB fields for the two gaps reduces the uncertainty and
yields B$_1$ = B$_2$ = (55 $\pm$ 20)~T for the a series. This value might also account for the data relevant to the 2a+$\delta$ series for which
MB fields between 30 T and 160 T are obtained. Nevertheless, a negative Dingle temperature is obtained assuming MB field in this range even
though the Dingle temperature for the a oscillation series is T$_D$(a) = 0.4~K. In addition, Dingle plot for the a+$\delta$ series can only be
accounted for by lower MB fields i. e. between 0.2 T and 19 T (above 19 T, negative T$_D$ values are obtained). Hence, the data in
Fig.~\ref{Fig8} cannot be accounted for by a unique set of MB field values. This may suggest that, in addition to SdH and QI, other contribution
should play a role in the oscillatory data.

\section{\label{Sum}Summary and conclusion}

The oscillatory behavior of the interlayer magnetoconductance of the quasi-two dimensional organic metal
(BEDT-TTF)$_8$Hg$_4$Cl$_{12}$(C$_6$H$_5$Cl)$_2$ can be described on the basis of linear combinations of three basic frequencies arising from the
compensated closed hole and electron orbits and from the two orbits located in between. It can be remarked first that the various MB-induced SdH
orbits and QI paths responsible for the observed oscillation spectrum are not independent but do constitute an interlinked network which has
been considered in the framework of the coupled orbits model of Falicov and Stachowiak \cite{4}. On the basis of the derived values of the
effective cyclotron masses linked to the various oscillation series, it can be inferred that a strong contribution of conventional SdH accounts
for the 2a+$\delta$ series while data for a+$\delta$ and b, are consistent with QI. Oppositely, the low values of m$_c(\delta)$ and
m$_c(3a+\delta$) disagree with both SdH and QI. In addition, the field dependence of the oscillation amplitude of the various series cannot be
consistently accounted for by a unique set of MB gaps E$_1$ and E$_2$. These features suggest that additional contribution, such as frequency
mixing due to oscillation of the chemical potential \cite{7,9} or interplay of electronic states from the different bands crossing the Fermi
level \cite{10} strongly influence the oscillatory behavior. From the experimental point of view, further enlightenment could be given by dHvA
experiments in high magnetic field. Indeed, contrary to conductivity, magnetization, as a thermodynamic parameter, is not sensitive to QI. In
addition, the configuration of measurement (in-plane vs. interlayer) should also be considered since, up to now, no frequency combinations has
been observed in conductivity data recorded in the in-plane configuration \cite{14}.

\begin{acknowledgments}
The authors would like to thank T. Ziman, J. Y. Fortin, R.
Fleckinger and V. Laukhin for discussions on chemical potential
oscillations and QI, E. Canadell for discussions about FS
calculations and J. Singleton for a very pertinent remark related
to the b oscillation series.

\end{acknowledgments}

$\dagger$ author for correspondence: audouard@insa-tlse.fr

\bibliography{apssamp}

\begin{references}
\bibitem{20} As already observed \cite{21}, even stronger downward curvatures and lower slopes of the Dingle plots are obtained in the three-dimensional case (n = 1/2 in Eq. (4)). This leads to lower MB fields and Dingle temperature e. g. $\sim$30~T and 0.2~K, respectively for the a oscillation series
\end{references}

\end{document}